\documentclass[
 reprint,amsmath,amssymb,
 aps, superscriptaddress, prl,
]{revtex4-2}

\usepackage{graphicx}
\usepackage{amsmath,amssymb}
\usepackage{orcidlink}
\usepackage{xcolor}
\usepackage{hyperref}
\hypersetup{
    colorlinks=true,
    linkcolor=blue,
    urlcolor=blue,
    citecolor=blue
    }

\begin{document}

\title{Optical response of the supersolid polaron}

\author{Laurent H. A. Simons \orcidlink{0009-0000-7251-5845}}\email{laurent.simons@uantwerpen.be}
\affiliation{
 Theory of Quantum and Complex Systems, Physics Department, Universiteit Antwerpen, B-2000 Antwerpen, Belgium
}

\author{Ralf Klemt \orcidlink{0009-0002-3231-447X}}
\affiliation{
 5. Physikalisches Institut and Center for Integrated Quantum Science and Technology,
Universit\"{a}t Stuttgart, Pfaffenwaldring 57, 70569 Stuttgart, Germany
}

\author{Tilman Pfau \orcidlink{0000-0003-3272-3468}}
\affiliation{
 5. Physikalisches Institut and Center for Integrated Quantum Science and Technology,
Universit\"{a}t Stuttgart, Pfaffenwaldring 57, 70569 Stuttgart, Germany
}

\author{Michiel Wouters \orcidlink{0000-0003-1988-4718}}
\affiliation{
 Theory of Quantum and Complex Systems, Physics Department, Universiteit Antwerpen, B-2000 Antwerpen, Belgium
}

\author{Jacques Tempere \orcidlink{0000-0001-8814-6837}}
\affiliation{
 Theory of Quantum and Complex Systems, Physics Department, Universiteit Antwerpen, B-2000 Antwerpen, Belgium
}

\date{\today}

\begin{abstract}

The ground-state properties of the supersolid polaron consisting of a neutral impurity immersed in a dipolar supersolid have recently been studied. Here, the optical response of an impurity in a dipolar supersolid is calculated and interpreted in terms of the contributions of the different excitation modes of the supersolid. The optical absorption spectrum reveals the two Van Hove singularities that correspond to the flattening of the two Goldstone modes of the supersolid at the Brillouin zone edge. A single peak is found in the superfluid regime corresponding to the roton minimum which diverges at the transition. We propose the response of an ionic impurity as an experimental probe for the supersolid excitations and show how this technique can be extended to neutral impurities with an electric or magnetic dipole moment.

\end{abstract}

\maketitle

\paragraph{Introduction --}

The polaron, as defined by Landau and Pekar \cite{landau1933electron,pekar1946local,landau1948effective}, refers to a charge in an ionic crystal which results in a lattice deformation that can be described by electron-phonon interactions \cite{alexandrov2010advances,devreese2016fr,*devreese2009frohlich,*devreese2006frohlich,cesare2021polarons}. The generalized polaron problem of a particle interacting with a bath via emission or absorption of bath excitations has since become one of the most basic but also important many-body problems \cite{alexandrov2010advances,cesare2021polarons,devreese2016fr,*devreese2009frohlich,*devreese2006frohlich}. Specifically, the Bose polaron, where an impurity is immersed in a (superfluid) Bose-Einstein condensate (BEC), has been the subject of many different theoretical \cite{tempere2009feynman,rath2013field,vlietinck2015diagrammatic,grusdt2015renormalization,ardila2015impurity,shchadilova2016polaronic,shchadilova2016quantum,levinsen2017finite,ichmoukhamedov2019feynman,ichmoukhamedov2022general} and experimental \cite{jorgensen2016observation,hu2016bose,yan2020bose,skou2022life} investigations due to the extreme controllability and tunability of ultracold atomic gases \cite{bloch2012quantum,bloch2008many}. It has since been extended to different configurations such as a charged impurity \cite{astrakharchik2021ionic,christensen2021charged,astrakharchik2023many,cavazos2024modified,simons2024path} or a rotor \cite{schmidt2015rotation,schmidt2016deformation} in a BEC and it has been probed using a variety of methods including radio-frequency spectroscopy \cite{jorgensen2016observation,hu2016bose,yan2020bose,skou2022life}, Bragg spectroscopy \cite{casteels2011response}, and absorption spectroscopy \cite{simons2024path}.

Recently, a supersolid phase has been discovered in a Bose-Einstein condensate consisting of dipolar atoms which exhibits periodic density modulations in addition to superfluidity \cite{tanzi2019observation,bottcher2019transient,chomaz2019long,norcia2021two}. The dipolar supersolid has a rich excitation spectrum containing multiple massless and massive modes \cite{guo2019low,natale2019excitation,tanzi2019supersolid,hertkorn2019fate,hertkorn2021density,hertkorn2024decoupled}. This is a very compelling phase to study the polaron problem in, as the impurity can be dressed by different excitation modes and reveal more interesting information about the excitations compared to the superfluid Bose polaron. The ground-state properties of such an impurity immersed in a one-dimensional dipolar supersolid such as the ground-state energy have been studied in Ref. \cite{simons2025polarons}.

In this Letter, the optical response of a charged impurity immersed in a one-dimensional dipolar supersolid is studied theoretically in the weak ion-dipole coupling regime within the Kubo formalism \cite{tempere2001optical,houtput2022optical}. We find that absorption spectroscopy of an ion in a dipolar supersolid can be used to probe the excitations of the supersolid as multiple peaks corresponding to different excitation modes can be observed in the spectra. In particular, the two Goldstone modes result in two visible Van Hove singularities in the absorption spectrum while the massive Higgs mode has a threshold frequency for absorption similar to the solid-state polaron \cite{devreese1971optical,devreese1972optical,mishchenko2000diagrammatic,tempere2001optical,houtput2022optical}. A single peak is found in the absorption spectrum in the superfluid regime due to the roton softening close to the transition. We also propose extensions to neutral dipolar impurities by using an oscillating magnetic field gradient instead of an oscillating electric field to probe the impurity.

\paragraph{Model --}

A model to describe an impurity immersed in a one-dimensional dipolar supersolid has been introduced in Ref. \cite{simons2025polarons} and will be used here to describe an ion in a dipolar supersolid. The one-dimensional dipolar supersolid is described using an effective one-dimensional extended Gross-Pitaevskii (eGP) theory where an anisotropic variational Gaussian ansatz is used for the transverse degrees of freedom $\psi_\perp(x, y)=\exp[-(\nu x^2+y^2/\nu)/(2(\sigma l_\perp)^2)]/(\sqrt{\pi}\sigma l_\perp)$ with $\sigma, \nu$ variational parameters and $l_\perp$ the harmonic oscillator length \cite{blakie2020variational,ilg2023ground}. Transverse polaron effects are neglected. The longitudinal direction can be described by the following periodic ansatz $\psi_0(z)=\sqrt{n_0}(1+\sum_{l=1}^\infty\Delta_l\cos(lk_sz))$, with $\Delta_l$ and $k_s$ being variational parameters and related to the periodic density modulation of the supersolid \cite{ilg2023ground}. The Bogoliubov approximation shifts the creation and annihilation operators with c-numbers $\Delta_l\sqrt{N_0}/2$ where $\Delta_l$ describes the strength of each contributing mode $l$ ($\Delta_0=2$) and $n_0=N_0/L$ the density of particles occupying the zero-momentum mode (this is different from the total condensate density $n$) \cite{ilg2023ground}. The variational parameters, and thus, the wavefunction of the supersolid can be found by minimizing the Gross-Pitaevskii energy including Lee-Huang-Yang corrections necessary for stability \cite{wachtler2016quantum,ferrier2016observation} via the local density approximation \cite{lima2011quantum,lima2012beyond,schutzhold2006mean, blakie2020variational,ilg2023ground}. An effective one-dimensional dipole-dipole interaction potential $V(q)$ is used by integrating out the transverse degrees of freedom using the aforementioned Gaussian ansatz \cite{blakie2020variational, ilg2023ground}. Here, an analytical approximation is used to simplify calculations \cite{blakie2020variational, ilg2023ground}. The excitation spectrum $\hbar\omega_l(q)$ and corresponding Bogoliubov coefficients $u^l_m(q), v^l_m(q)$ can be obtained by expanding and diagonalizing the supersolid Hamiltonian up to second order in creation and annihilation operators similar to Ref. \cite{ilg2023ground}. As expected, the resulting excitation spectrum has two Goldstone modes and a massive Higgs mode \cite{guo2019low,natale2019excitation,tanzi2019supersolid,hertkorn2019fate,hertkorn2021density,hertkorn2024decoupled}.

A Hamiltonian can be derived for a general impurity in a dipolar supersolid described by the aforementioned Bogoliubov theory similar to the superfluid Bose polaron \cite{tempere2009feynman,shchadilova2016quantum}. For weak ion-dipole coupling where beyond-Fr\"ohlich terms describing absorption or emission of multiple excitations by the impurity can be neglected, the Bogoliubov-Fr\"ohlich Hamiltonian describing an impurity with mass $m_I$ in a one-dimensional dipolar supersolid is given by \cite{simons2025polarons}
\begin{widetext}
\begin{equation}
\begin{aligned}
  \hat{H} &= \frac{\hat{p}_I^2}{2m_I}+\sum_{q,l}\hbar\omega_l(q)\hat{\alpha}^\dagger_{q,l}\hat{\alpha}_{q,l}+U(\hat{z}_I)+\sum_{q,l}V_l(q,\hat{z_I})\left(\hat{\alpha}_{q,l}+\hat{\alpha}^\dagger_{-q,l}\right),\\
  V_l(q, \hat{z}_I) &= \frac{\sqrt{N_0}}{2L}\sum_{m,m'}V_{\text{IB}}(q+mk_s)e^{i(q+mk_s)\hat{z}_I}\Delta_{m'}\left(u^{l}_{m-m'}(q)+v^{l}_{m-m'}(q)\right),\\
  U(\hat{z}_I)&=\frac{n_0}{4}\sum_{l,l'}V_{\text{IB}}(lk_s)\Delta_{l'-l}\Delta_{l'}e^{ilk_s\hat{z}_I}.
\end{aligned}
\end{equation}
\end{widetext}
Here, $\hat{\alpha}^\dagger_{q,l}, \hat{\alpha}_{q,l}$ are the creation and annihilation operators of an excitation mode $l$ with momentum $q$. The momenta $q$ in the summations are in the first Brillouin zone defined from $-k_s/2$ to $k_s/2$ and the index $m$ denotes the zone number. The above Hamiltonian can be interpreted as follows: an ion in a one-dimensional supersolid feels an interaction-dependent periodic background potential $U(\hat{z}_I)$ and can absorb or emit an excitation mode $l$ with momentum $q$. The latter interaction is described by the vertex factor $V_l(q, \hat{z}_I)$. Similar to the effective one-dimensional dipole-dipole interaction potential, an effective one-dimensional ion-dipole interaction potential $V_{\text{IB}}(q)$ is derived by integrating out the transverse degrees of freedom by using the Gaussian ansatz for the condensate and a similar wavefunction but with $\sigma=\sqrt{m_B/m_I}$ and $\nu=1$ for the impurity, see Ref. \cite{simons2025polarons}. The three-dimensional ion-dipole interaction potential used is approximated by the ion-atom interaction potential used in the literature for the charged superfluid Bose polaron \cite{astrakharchik2021ionic,christensen2021charged,astrakharchik2023many,cavazos2024modified,simons2024path} given by \cite{krych2015description}
\begin{equation}
    V^{\text{(3D)}}_{\text{IB}}(r)=-\frac{C_4}{(r^2+b^2)^2}\frac{r^2-c^2}{r^2+c^2},
\end{equation}
where $C_4$ is related to the polarizability of the atoms, $b$ is related to the depth of the potential, and $c$ creates a repulsive barrier.
Similar to Ref. \cite{simons2025polarons}, the periodic potential $U(\hat{z}_I)$ is taken into account by introducing an effective Bloch mass $m^*$ obtained by calculating the Bloch spectrum of the ion in the periodic potential which replaces the bare impurity mass. We restrict the discussion to the lowest Bloch band and therefore require the distance between the first and second Bloch bands to be larger than the largest energy considered within our theory, which is given by $E_\perp=\hbar^2/(m_Bl_\perp^2)$.
In addition, the values are chosen such that the polaron is delocalized over many droplets (weak ion-dipole coupling) so that our theory and the approximations that have been made are valid.

\paragraph{Optical response --}

The optical response of the ion is studied by employing linear response theory and the Kubo formalism, following Refs. \cite{tempere2001optical,houtput2022optical, simons2024path}. The memory function formalism used there is valid in the weak-coupling regime as the expectation values are factorized such that they can be calculated analytically. The memory function expression used for the optical absorption $\Lambda(\omega)$ can be simplified resulting in
\begin{multline}
    \Lambda(\omega)\epsilon_0cn_r\mathcal{V}/e^2=\frac{n_0}{4(m^*)^2\hbar\omega^3}\sum_{l,m,s}\Delta_m\Delta_s\int\limits_0^{k_s/2}dk k^2V^2_{\text{IB}}(k)\\(u^{l}_{m}(k)+v^{l}_{m}(k))(u^{l}_{s}(k)+v^{l}_{s}(k))\delta\left(\omega-\frac{\hbar k^2}{2m^*}-\omega_l(k)\right),
\end{multline}
with $\epsilon_0$ the vacuum permittivity, $c$ the speed of light, $n_r$ the refractive index, $e$ the charge of the ion, and $\mathcal{V}$ the volume of the system.
The zero-temperature limit is studied here and, as there is only a single ion, the delta function has been used for the dynamical structure factor of the impurity. In the superfluid regime, the indices are dropped in the above equation (with $\Delta_l=2\delta_{l,0}$) and the integral is performed from zero to infinity instead of the Brillouin zone edge $k_s/2$.

\begin{figure}
    \centering
    \includegraphics[scale=1]{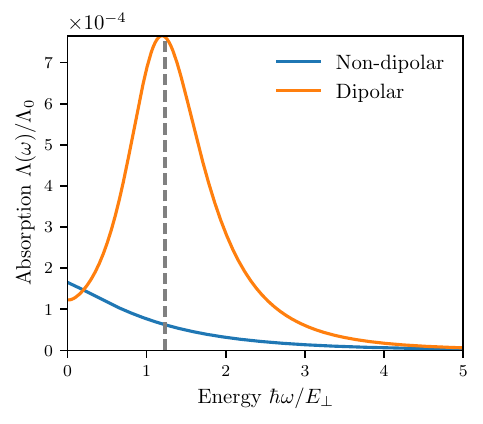}
    \caption{Optical absorption of an ion in a one-dimensional superfluid with contact interactions ($\epsilon_{\text{dd}}=0$) or dipolar interactions ($\epsilon_{\text{dd}}=1.2$). The dashed grey line indicates the energy $\hbar^2k_r^2/(2m_I)$ with $k_r$ the roton momentum. The following values are used: $C_4=0.01\hbar^2l_\perp^2/m_B$, $b=\l_\perp$, $c=0.00023l_\perp$, $m_I=m_B$, $n=2386.2/l_\perp$, and $a_s=0.005l_\perp$.}
    \label{fig1}
\end{figure}

The optical absorption spectrum of an ion in a one-dimensional superfluid is shown in Figure \ref{fig1} with the unit $\Lambda_0=e^2l_\perp^2/(\epsilon_0cn_r\hbar \mathcal{V})$. The values used are $C_4=0.01\hbar^2l_\perp^2/m_B$, $b=\l_\perp$, $c=0.00023l_\perp$, $m_I=m_B$, $n=2386.2/l_\perp$, $a_s=0.005l_\perp$, and $\epsilon_{\text{dd}}=0$ (non-dipolar) or $1.2$ (dipolar), taken such that the ion-dipole coupling is weak. The values of $C_4$, $b$ and $c$ used in the ion-dipole interaction potential are similar to Refs. \cite{massignan2005static,christensen2021charged,cavazos2024modified,simons2024path} where Rubidium-87 atoms are used for the condensate. Notice that the absorption goes to a finite value for small frequencies compared to the three-dimensional superfluid Bose polaron and decays for larger frequencies. A discussion about possible delta peaks needed to satisfy the conductivity sum rules is out of the scope of this paper, similar to Ref. \cite{simons2024path}. The spectrum for the dipolar superfluid shows a single peak that becomes larger as a function of $\epsilon_{\text{dd}}$ until it diverges at the transition $\epsilon_{\text{dd, c}}=1.34$. The position of the peak is close to $\hbar^2k_r^2/(2m_I)$, the energy corresponding to an impurity with the roton momentum $k_r$ at the transition \footnote{The peak is situated outside of the regime of validity of our theory $\hbar\omega<E_\perp$. However, we expect that transverse effects will not change the peak qualitatively. In addition, note that there is no analytical expression for the peak position before the transition.}. Eventually, at the transition point, the position of the divergent peak coincides with $\hbar^2k_r^2/(2m_I)$.

\begin{figure}
    \centering
    \includegraphics[scale=1]{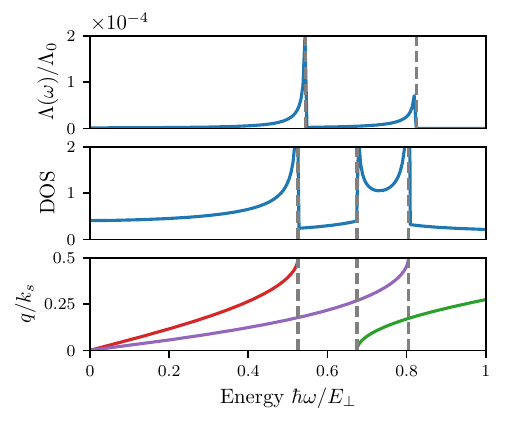}
    \caption{Optical absorption of an ion in a one-dimensional supersolid, together with the supersolid density of states and excitation spectrum. The dashed grey lines on the middle and bottom panel indicate the energies where the excitation spectrum becomes flat, while for the top panel the dashed grey lines show the energies where the Goldstone modes become flat plus the kinetic energy $\hbar^2(k_s/2)^2/(2m^*)$ of an impurity with momentum $k_s/2$. The following values are used: $C_4=0.01\hbar^2l_\perp^2/m_B$, $b=\l_\perp$, $c=0.00023l_\perp$, $m_I=m_B$, $n=2386.2/l_\perp$, $a_s=0.005l_\perp$, and $\epsilon_{\text{dd}}=1.345$.}
    \label{fig2}
\end{figure}

In the supersolid regime, the optical response changes drastically. Two peaks are readily observed in the absorption spectrum which diverge at a certain value of the energy. To gain more intuition about the peaks, the density of states given by $\text{DOS}=\sum_l\int_{\text{BZ}}dq \delta(\hbar\omega-\hbar\omega_{q,l})/(2\pi)$ is calculated. The absorption spectrum, density of states, and excitation spectrum are plotted in Figure \ref{fig2} for the supersolid regime using the same values as before but with $\epsilon_{\text{dd}}=1.345$. In the density of states, three Van Hove singularities can be observed corresponding to energies where the excitation bands flatten (at momentum $k_s/2$ for the Goldstone modes, while for the Higgs mode at momentum zero). The two van Hove singularities which correspond to the Goldstone modes are also observable in the absorption spectrum. The positions of the two diverging peaks correspond to the energies where the Goldstone modes flatten plus an additional kinetic energy $\hbar^2(k_s/2)^2/(2m^*)$ for the impurity (the grey dashed lines). Experimentally, these peaks will be finite and broadened by other effects, such as finite frequency sampling, temperature broadening, and the finite lifetime of the system. Nevertheless, the peak positions can be used to infer information about the excitation spectrum such as the energies of the Goldstone modes at the Brillouin zone edge, which can be used to infer the superfluid fraction \cite{vsindik2024sound,biagioni2024measurement,platt2025supersolid}. Also note that the absorption spectrum is plotted for values of the energy up to $E_\perp$ as transverse excitations may play a role when $\hbar\omega>E_\perp$ which are not taken into account within our theory.

\begin{figure}
    \centering
    \includegraphics[scale=1]{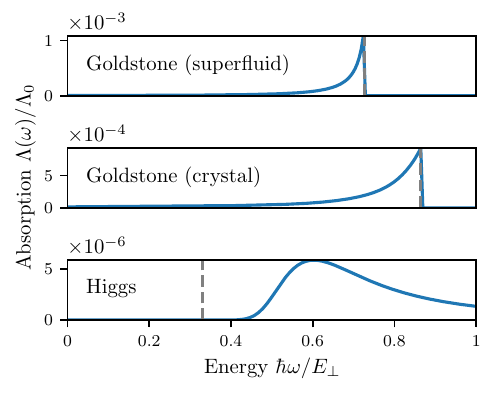}
    \caption{Optical absorption of an ion in a one-dimensional supersolid showcasing the contribution of the different excitation modes. The dashed grey line on the bottom panel shows the Higgs mode energy for momentum zero. The following values are used: $C_4=0.01\hbar^2l_\perp^2/m_B$, $b=\l_\perp$, $c=0.00023l_\perp$, $m_I=m_B$, $n=2386.2/l_\perp$, $a_s=0.005l_\perp$, and $\epsilon_{\text{dd}}=1.341$.}
    \label{fig3}
\end{figure}

However, the Higgs mode Van Hove singularity is not immediately visible in the absorption spectrum. To study the absorption spectrum in more detail, Figure \ref{fig3} shows the contribution of the main excitation modes to the absorption spectrum for the same values as Figure \ref{fig2} (except $\epsilon_{\text{dd}}=1.341$ to show the Higgs contribution more clearly by lowering the Higgs band into the energy range where our theory is valid). It can be seen that the Higgs mode has a contribution to the absorption spectrum which is much smaller than the Goldstone modes. The contribution is also a single peak similar to the three-dimensional superfluid Bose polaron \cite{simons2024path}.
The reason why the Van Hove singularity is not pronounced for the Higgs mode in the absorption spectrum is because the rest of the absorption equation excluding the delta function goes to zero faster than the divergence in the density of states. The absorption starts at energy $\hbar\omega=\hbar\omega_{-1}(0)$, the energy of the Higgs mode at momentum zero, similar to the solid-state polaron absorption.

\begin{figure}
    \centering
    \includegraphics[scale=1]{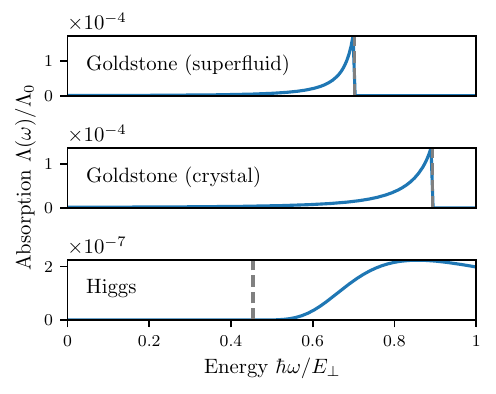}
    \caption{Optical absorption of a dipolar impurity in a one-dimensional supersolid using the same equation as the ionic impurity but with a different interaction potential to get a feel of the actual impurity response. The dashed grey line on the bottom panel shows the Higgs mode energy for momentum zero. The following values are used: $a_s^{(\text{IB})}=0.001l_\perp$, $\epsilon_{\text{dd}}^{(\text{IB})}=0.1$, $m_I=m_B$, $n=2386.2/l_\perp$, $a_s=0.005l_\perp$, and $\epsilon_{\text{dd}}=1.342$.}
    \label{fig4}
\end{figure}

\paragraph{Extension to neutral impurities --}
The idea of optical absorption spectroscopy requires a charged impurity which can respond to the applied electric field. However, the above results could potentially be observed even when using neutral impurities which are much easier to create and control experimentally. An example of a spectroscopy method similar to absorption spectroscopy used to study the neutral impurity is Bragg spectroscopy \cite{casteels2011response}, however it is quite different from absorption spectroscopy as it also resolves momentum. Experimentally, instead of ionic impurities also neutral impurities are a promising candidate to measure the absorption spectra. If these impurities possess a magnetic or electric dipole moment,
they can respond to an oscillating magnetic/electric field gradient
similarly as charged impurities to an AC electric field. As the supersolid is made up of either magnetic or electric dipoles ideally the impurity has the other dipole moment, such that the AC modulation of the gradient field only affects the impurity and not the supersolid itself. To probe the similarity between the two systems, Figure \ref{fig4} shows the absorption spectrum of a dipolar impurity immersed in the supersolid where the same equation as the ionic impurity is used but with a different interaction potential $V_{\text{IB}}(q)$ where $V^{(\text{3D})}_{\text{IB}}(q)=2\pi\hbar^2a_s^{(\text{IB})}/m_r\left(1+\epsilon_{\text{dd}}^{(\text{IB})}(3q_y^2/q^2-1)\right)$ \cite{simons2025polarons, ardila2018ground}. This is done to see if the interaction potential (dipolar-dipolar or ion-dipolar) has a significant effect on the observed spectrum. Here, $a_s^{(\text{IB})}$ is the s-wave impurity-dipole scattering length, $m_r=m_Im_B/(m_I+m_B)$ is the reduced mass, and $\epsilon_{\text{dd}}^{(\text{IB})}=\mu_0\mu_I\mu_Bm_r/(6\pi\hbar^2a^{(IB)}_s)$ with $\mu_I$ the impurity magnetic dipole moment \cite{simons2025polarons, ardila2018ground}. The values used are $a_s^{(\text{IB})}=0.001l_\perp$, $\epsilon_{\text{dd}}^{(\text{IB})}=0.1$, $m_I=m_B$, $n=2386.2/l_\perp$, $a_s=0.005l_\perp$, and $\epsilon_{\text{dd}}=1.342$. It can be seen that the features observed in the absorption spectrum of an ionic impurity remain qualitatively the same when using the dipolar interaction potential. This shows that neutral dipolar impurities can also be used as an experimental probe for the supersolid excitations.

\paragraph{Summary and outlook --} A probing method has been introduced where an ionic impurity immersed in a dipolar supersolid can reveal information about the supersolid excitation spectrum using absorption spectroscopy. Specifically, two Van Hove singularities corresponding to the Goldstone modes are observed in the density of states and absorption spectrum. The Higgs mode results in a much smaller single peak starting at a minimum frequency corresponding to the creation of a single Higgs excitation at momentum zero. A precursor peak to the Van Hove singularities can be found in the superfluid regime linked to the roton minimum which diverges at the transition. Finally, a possible extension to neutral dipolar impurities is proposed by using an oscillating magnetic field gradient instead of an AC electric field. The next step would be to observe the optical response experimentally for a charged impurity and test our possible extension to dipolar impurities. The effect of trapping potentials and different geometries may impact the response in an interesting way as, in our theory, an infinitely long 1D supersolid is used. Another compelling future direction is to study the behavior of the optical response of a localized ion in the strong-coupling regime \cite{simons2025polarons} similar to the Holstein polaron, see e.g. \cite{goodvin2011optical}. It would also be interesting to study these impurity systems in other supersolids, such as a spin-orbit-coupled condensate in the supersolid phase \cite{li2017stripe}, where the supersolidity arises from other types of interactions.

\paragraph{Acknowledgments --}
This work is supported financially by the Research Foundation–Flanders (FWO), Projects numbers G0A9F25N, G0AIY25N, GOH1122N, G061820N, and G060820N, and by the University Research Fund (BOF) of the University of Antwerp. R.K. and T.P. acknowledge funding from the European Research Council (ERC) (Grant Agreement No. 101019739).

\end{document}